\documentclass[preprint,showpacs,preprintnumbers,amsmath,amssymb]{revtex4-1}

\usepackage[]{graphicx}
\usepackage{dcolumn}
\usepackage{bm}

\begin{document} 
\title{A scaling-limit approach to the theory of laser transition} 
 
\author{Paul Gartner}
\affiliation{Institute for Theoretical Physics,
        University of Bremen, 28334 Bremen, Germany \\
         and \\
         National Institute of Materials Physics,
        Bucharest-M\u agurele, Romania}

\email{gartner@itp.uni-bremen.de}

\date{\today}
  
\begin{abstract}
The conditions for the appearance of a sharp laser transition are
formulated in terms of a scaling limit, involving vanishing cavity
loss and light-matter coupling, $\kappa \to 0$, $g \to 0$,
such that $g^2/\kappa$ stays finite. It is shown analytically that in
this asymptotic parameter domain, and for pump rates above the threshold
value, the photon output becomes large in a sense that is specified,
and the photon statistics becomes strictly Poissonian. Numerical
examples for the case of a two-level and a three-level emitter are
presented and discussed in relation to the analytic result.   
\end{abstract}

\maketitle
\section{INTRODUCTION} \label{sec:intro} 

An early preoccupation of laser theory was the analogy between the
onset of lasing and phase transitions\cite{deG,haken,richter}. Within the
cavity QED models the problem became increasingly accessible to accurate
numerical treatments, and evidence that an abrupt change of regime
takes place has accumulated. In parallel, approximate analytic
considerations also argued in the same sense. 

A particularly illuminating case is the so-called random injection
model of Scully and Lamb\cite{scullylamb}, which has the advantage of
addressing directly the cavity mode statistics. The model was
extensively studied \cite{stenh} and is by now textbook
material\cite{WallsMilb,Orszag}. 
In emitter-plus-mode models the situation is more complicated because
the photonic state information has first to be extracted, by
eliminating the emitter degrees of freedom. This procedure can be carried
out\cite{agarwal} in the case of single emitters with few (usually
only two) states. Then information about the photon statistics can be
obtained, either from the numerics or, using simplifying
approximations, analytically too\cite{karlkil,boozer, ricecarm,
delvferm,delvstrong, auff}.
 
In this context the seminal paper by Rice and Carmichael, Ref.
\citenum{ricecarm}, has drawn the attention on the necessity of a
limiting procedure for obtaining a sharp transition, with a precisely
defined threshold. The analogy with the thermodynamic limit in the
theory of phase transition was invoked. It was argued that for the
lasing transition 
the limit involves the $\beta$-factor going to 0, together with the
cavity loss $\kappa$ so that the ratio $\beta/\kappa$ remains
finite. In this limit, and for a pump rate exceeding a threshold value, 
the number of photons $N$ becomes infinite, generating ``an explosion
of stimulated emision''\cite{ricecarm}, and the appropriate object of
study is the rescaled value $\beta N$, which stays finite.  
   
In the present paper we show that a similar scaling limit is needed
for an abrupt onset of lasing, but we use in our formulation the more
ubiquitous Jaynes-Cummings (JC) coupling constant $g$, instead of the
$\beta$-factor. The latter is proportional to $g^2$ and indeed our
scaling procedure implies $\kappa \to 0$, $g \to 0$ so that
$g^2/\kappa$ is finite. In this scaling limit and above the threshold
one obtains $N \to \infty$ but such that $\kappa N$ remains
finite. Moreover we are able to prove not only that the transition
becomes abrupt, but also that the photon statistics above the
threshold turns exactly Poissonian. The proof is analytic and does not
rely on approximations. Also, analytic expression for relevant data,
like the threshold pump rate, level occupancies and photon output in
the lasing regime, are obtained. Numerical results are shown as
illustration of the statements.   
    
\section{THE MODEL AND STATEMENT OF THE RESULT} \label{sec:model}

Single emitter lasers are commonly described as embedded
JC systems. In other words, two emitter states, either quantum dot or
atomic configurations, interact with the cavity mode via the JC
Hamiltonian
\begin{equation}
H_{JC} = g\, b^{\dagger}\left|2\right>\left<1\right| + g\, b
\left|1\right>\left<2\right| \; ,
\label{eq:jc}
\end{equation}
in the presence of, possibly, other states. Here $b,b^{\dagger}$ are
the photonic operators and the emitter states are denoted by
$\left|i\right>$. In particular $\left|1\right>$ and $\left|2\right>$
specify the upper and the lower laser state, respectively. We assume
that the cavity mode is resonant with the laser transition. 
Dissipation effects are included in the master equation for the
density operator $\rho$ (in the interaction picture and with
$\hbar=1$)
\begin{equation} 
\frac{\partial}{\partial t} \rho = -i \left[H_{JC}, \rho \right] +
{\cal{L}}(\rho) 
\label{eq:master}
\end{equation}   
by the Lindblad terms  
\begin{equation}    
{\cal{L}}(\rho) = \frac{\kappa}{2} \left[2b \rho b^{\dagger} - 
  b^{\dagger}b \rho -\rho b^{\dagger}b \right] + 
  \sum_{(i,j)}  \frac{\gamma_{ij}}{2} \left[2\sigma_{ij} \rho
   \sigma_{ij} ^{\dagger} - \sigma_{ij}^{\dagger}\sigma_{ij} \rho -\rho
  \sigma_{ij}^{\dagger}\sigma_{ij} \right] \; , \quad
  \sigma_{ij} = \left|i\right> \left< j \right| \; .
\label{eq:lindblad}
\end{equation} 
A central role here is played by the first term, describing the cavity losses
at a rate $\kappa$, while the second term summarizes transition processes from
states $\left|j\right>$ to $\left|i\right>$ at rates $\gamma_{ij}$. Lowering 
$\sigma_{ij}$ operators and their hermitian conjugates $\sigma_{ij}
^{\dagger}$ give rise to Lindblad terms accounting for relaxation,
but raising terms are considered as well, to simulate incoherent pumping
\footnote{Coherent excitation can be described by an
  additional Hamiltonian term\cite{musav}, but we will not discuss
  this approach here.}, and the corresponding rate will then be denoted by $P$. 

The master equation is solved in time until a steady-state solution is
reached, from which data concerning level occupancies and photon
statistics can be extracted, as function of the pumping rate and other
parameters. With this information at hand one can address the problem
of laser transition: how to identify it and what are the conditions
for its appearance. 

The answer to the first question is that the lasing regime is defined
by accumulation of a large number of photons in the cavity, the
statistics of their number $n$ obeying a Poissonian law
\begin{equation} 
\rho_{n,n} = \sum_i \rho^{i,i}_{n,n} = \frac{\lambda^n}{n!}
e^{-\lambda} \quad , \quad \mathrm{for \, all \;\;} n \;   .
\label{eq:poisson}
\end{equation}       
Equivalently, the Poisson statistics amounts to the requirement that
the normal-ordered expectation values $p_n= \left< 
b^{\dagger \,n}b^n \right>$  depend exponentially on $n$, $p_n =
\lambda^n$, or that the zero time-delay $n$-th order correlation functions
$g^{(n)}=p_n/p_1^n$ all become equal to 1. The problem is that,
on the one hand, the  large number condition is imprecise (how large
is large?) and, on the other hand, no matter which of these 
criteria for Poisson statistics one chooses to apply, one has to check
an infinite set of equalities. This is practically impossible, 
either experimentally or numerically. This is why it is often encountered in
the literature that one limits oneself to simpler lasing
criteria, like  $N >1$ and $g^{(2)}=1$. 
       
The aim of the present paper is to formulate and prove the conditions
under which strict Poissonian statistics is generated and at the same
time to specify in what sense the photon output becomes large. Essentially
these conditions involve the limit $\kappa \to 0$ and simultaneously 
$g \to 0$, but with the JC coupling parameter scaling like $\sqrt{\kappa}$
so that the ratio $g^2/\kappa$ remains finite. The necessity of a certain
limiting procedure for obtaining a well-defined transition to a purely
Poissonian statistics was recognized long ago\cite{ricecarm} and is
analogous to the thermodynamic limit in the theory of phase transitions.
The precise formulation of our statement runs as follows:

{\em
a. Rescale $\kappa$ as $\varepsilon\, \kappa$ and $g$ as
$\sqrt{\varepsilon} g$. Then, in the limit $\varepsilon \to 0$ and for
the pump rate above a threshold value, the average photon number 
$N=p_1$ tends to infinity in such a way that the product $\varepsilon N$ 
remains finite.

b. In the same limit and above the threshold, the rescaled
expectation values $\tilde p_n= \varepsilon^n\, p_n$ remain finite for
all $n$, with their limit values obeying an exponential law  $\tilde
p_n= \tilde p_1^n$. Equivalently, all the correlation functions become
$g^{(n)}=1$. Below the threshold the increase in photon production is
not sufficient, leading to $\tilde p_n =\varepsilon^n\, p_n \to 0$ for
all $n \geq 1$, and the correlation functions are in general different
from 1.
 
c. In this scaling limit the transition between the two regimes
becomes sharp, with a well-defined threshold point.   
} 

Several comments are in order. The model parameters, $g$ and $\kappa$,that 
are rescaled correspond to the photon source and sink, respectively. They are
present in all laser models and therefore the formulation of the scaling limit
in these terms makes sense in various situations. The limit of small cavity
losses (good cavity) pleads in favour of photon accumulation, but with the
simultaneous diminishing of the production rate, proportional to $g^2$, their
overall increase is not a foregone conclusion. The statement (a) above spells
out what is meant by a large photon number, namely that it should scale like
$1/\kappa$ to compensate for the reduction of the rate of escape from
the cavity. Thus, even for a cavity quality factor $Q$ close to infinity there
is still light coming out from the device.  

In situations as those described by incoherent excitation, at high pump
rates the phenomenon of self-quenching\cite{musav} might
occur. Beside producing population inversion, the pump can destroy
coherence between the laser levels and inhibit the transition. This
plays against lasing and therefore one may encounter a double transition,
one at the onset of lasing and the other when lasing becomes quenched.
In such cases the lasing regime takes place in a given interval of the
excitation rates, limited below by the threshold value and above by the
self-quenching $P_{thr}<P<P_{sq}$. The endpoints of this interval
depend on the model parameters. Accordingly, it may occur that the
interval shrinks to zero and then no transition takes place. Such
situations will also be discussed below. 

In what follows we will first bring numerical evidence in favor of the
scaling limit result. We illustrate the situation with calculations
performed on a two-level and on a three-level model. In both cases the
scaling limit tendencies are quite clear. Still, numerical statements
do not amount to a proof, which can only rely on an analytic
argument. We are able to formulate an analytic derivation of the
scaling limit result in the two-level case (see 
Sec.~\ref{sec:tla}). Also, an analytic proof is available\cite{pg} for the
random injection model\cite{scullylamb,stenh}, which does not belong
to the class defined by Eqs.~(\ref{eq:master}),(\ref{eq:lindblad}). We
believe that all these results speak in favor of a wider generality of the
scaling limit statement. 
    
\subsection{Numerical results} 
\label{subsec:numerical} 

\begin{figure}
   \begin{center}
   \includegraphics[height=5cm]{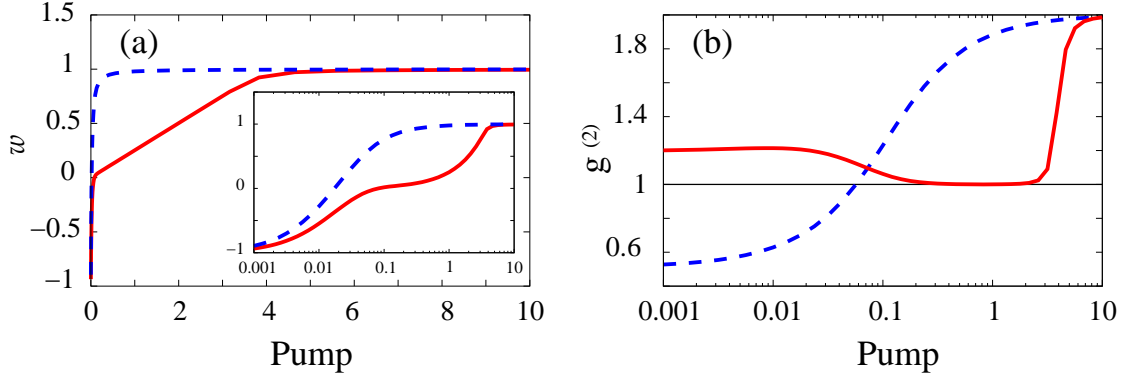}
   \end{center}
   \caption
   { \label{fig:one}
 (a) Population inversion $w$  in linear and (inset) semilogarithmic
     plot, and (b) second order correlation function 
 $g^{(2)}$ for a two-level model. The parameters are, red (solid)
 line: $\gamma=0.02$,  $g=0.1$ and $\kappa= 0.01$, blue (dotted) line:
 $\gamma=0.01$,  $g=0.01$ and $\kappa= 0.02$. 
\vspace{0.3cm} 
 }
   \end{figure}

Steady-state results are obtained from the long-time limit of the
master evolution. Particularly indicative of a transition is the
behavior of the population inversion  $w = \left< \, \left| 1
\right> \left< 1 \right|- \left|2 \right> \left< 2 \right|  \,
\right> $ as a function of the pump. Depending on the parameters, 
one clearly detects two types of behavior\cite{delvferm,auff} 
illustrated in Fig.\ref{fig:one}(a) \footnote{The steady-state
  results depend only on the ratio of the parameters, therefore their
  units are irrelevant and not specified.}. 
The plots correspond to a two-level  
emitter in which the pumping is described by the raising
$\left|2\right> \to \left|1\right>$ Lindblad operator with the rate 
$P = \gamma_{12}$, while for the rate of loss to non-lasing modes we
use the notation $\gamma_{21}=\gamma$. For one set of parameters the
curve is strictly concave, while for the other there appears an almost
perfectly linear shortcut\cite{karlkil, delvmoll, pg} separating two
concave regions. This clearly suggests that in the latter case an
abrupt change of regime is taking place, in a given $P$-interval. It
is also obvious that the appearance of 
the transition is conditioned by certain parameter values. By
examining Fig.\ref{fig:one}(b), we see that this linear segment (which
in a semilogarithmic representation appears as an additional convex
region, see  inset of Fig.\ref{fig:one}a),
corresponds to $g^{(2)}$ being very close to unity, which is
characteristic for coherent light. Therefore it is natural to assume
that we are in the presence of the lasing regime. Moreover, the
fact that for large $P$-values the linear behavior disappears is
consistent with the inhibition of lasing by self-quenching.

\begin{figure}
   \begin{center}
   \includegraphics[height=5cm]{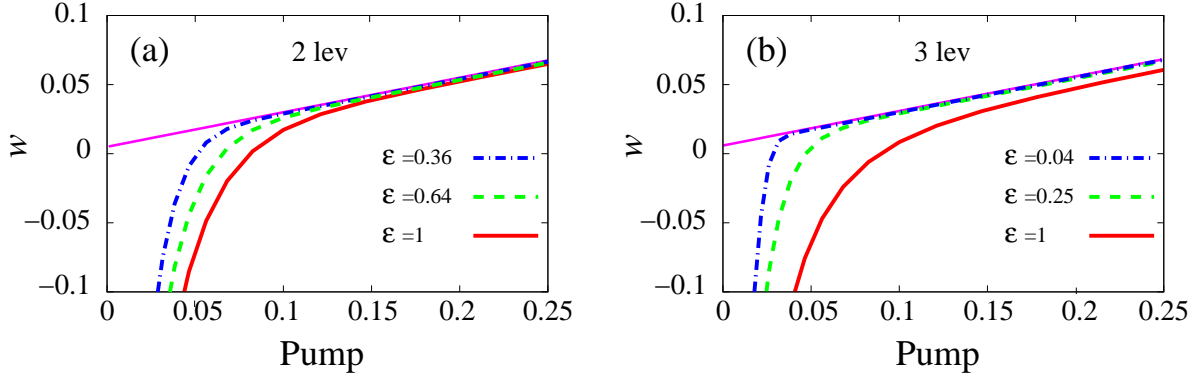}
   \end{center}
   \caption
   { \label{fig:two}
 Scaling parameter dependence of the population inversion $w$ for (a)
 the two-level emitter with parameters: $\gamma=0.02$,
 $g=\sqrt{\varepsilon}\, 0.1$  and $\kappa= \varepsilon \, 0.01$ and
 (b) the three-level emitter with  parameters: 
  $\gamma=0.02$, $\gamma'=0.05$, $g=\sqrt{\varepsilon} \, 0.1$ and 
  $\kappa= \varepsilon\, 0.01$. Thin solid lines
 correspond to the analytic result Eq.~(\ref{eq:scaled_ansatz_w}). 
\vspace{0.3cm} 
 }
   \end{figure}

Zooming in on the leftmost point of the linear interval, as in
Fig.\ref{fig:two}, it is seen that the transition becomes more and
more abrupt as the scaling parameter becomes smaller, in 
accordance with the scaling limit statement. 
This is seen both in the panel (a) of Fig.\ref{fig:two}, which refers to the
two-level model discussed in Fig.\ref{fig:one}, 
and in the panel (b), which shows the result for a three-level emitter. In this
latter case the pump is raising the system from the lower laser state
$\left|2\right>$ to a third state $\left|3\right>$ ($P = \gamma_{32}$),
wherefrom it relaxes to the upper laser state $\left|1\right>$ with the rate
$\gamma'=\gamma_{13}$. As before $\gamma_{21}=\gamma$.

   \begin{figure}
   \begin{center}
   \includegraphics[height=5cm]{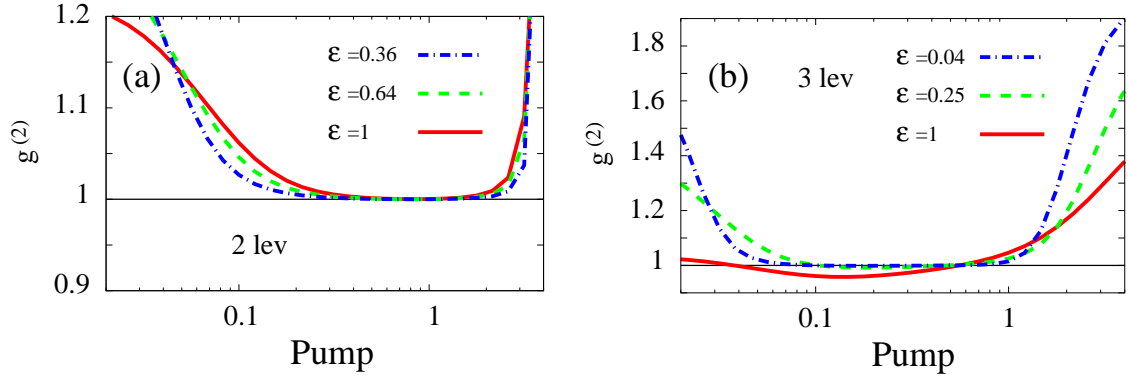}
   \end{center}
   \caption
   { \label{fig:three}
 Scaling parameter dependence of $g^{(2)}$ for the same parameters as
 in Fig.\ref{fig:two}.
\vspace{0.5cm} 
 }
   \end{figure} 
The same tendency is seen in plots of $g^{(2)}$, which becomes closer
and closer to the coherent light value $g^{(2)}=1$, as the scaling
parameter becomes smaller, both for the two- and for the three-level
case, as illustrated in Fig.\ref{fig:three}.

\begin{figure}
   \begin{center}
   \includegraphics[height=5cm]{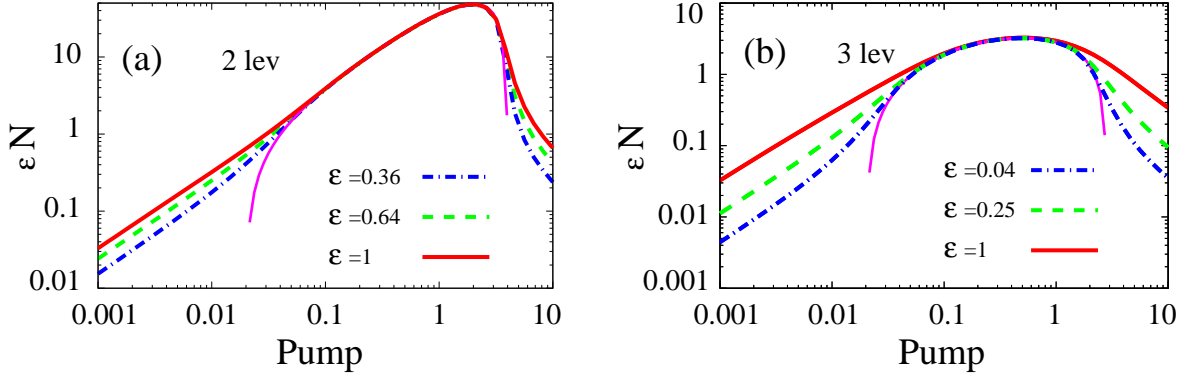}
   \end{center}
   \caption
   { \label{fig:four}
 Rescaled photon output for the two- and three-level emitter
 cases. The parameters are the same as in
 Fig.\ref{fig:two}. Thin solid lines correspond to
 Eq.~(\ref{eq:scaled_ansatz_N}) in panel (a),  and to
 Eq.~(\ref{eq:scaled_ansatz_N_3}) in panel (b). 
 }
   \end{figure}

Finally, in Fig.\ref{fig:four} we show numerical results for the
photon output. It is seen that in the lasing interval the rescaled photon numbers $\tilde N =
\varepsilon N$ have practically reached their limit values, given by
Eqs.~(\ref{eq:scaled_ansatz_N}) and~(\ref{eq:scaled_ansatz_N_3}), also plotted
in the figure. This means that indeed, $N$ grows like $1/\varepsilon$, in
accordance with the scaling statement.

The agreement seen in this section between the numerical data and the scaling
limit results can hardly be accidental. Values of $g^{(2)}$ close to unity
suggest that in the interval of intermediate pump strengths the system operates
in the lasing regime, and the large photon output speaks in favor of
this supposition too. Nevertheless, it is not at all clear yet how the linear
dependence of the population inversion on the excitation is in any way linked to
lasing. The analytic arguments of the next section will prove   
that, indeed, the two are related and appear simultaneously. 

\section{THE TWO LEVEL LASER}\label{sec:tla} 
 
A special feature of Eq.~(\ref{eq:master}) is the fact that
it provides a system of closed equations for a subclass of relevant
density matrix elements. In this category, the elements which are
diagonal in the emitter states $ \left|i\right>$ are also diagonal in the
photon number $n$, $\rho^{i,i}_{n,n}$, and the only off-diagonal elements
are of the form  $\rho^{1,2}_{n,n+1}$ and their complex conjugates
$\rho^{2,1}_{n+1,n}$. This is a consequence of fact that the JC
Hamiltonian conserves the excitation number $\left|1\right> \left< 1
\right| +  b^{\dagger}b$. 

We consider here the case of an emitter consisting of only the two
laser states $\left|1\right>$ and $\left|2\right>$. The Lindblad
terms describe, beside the cavity losses, the spontaneous emission
into non-lasing modes and the pumping, with the  rates
$\gamma=\gamma_{21}$ and $P=\gamma_{12}$, respectively. The master
equation implies an infinite set of equations of motion for
expectation values. The above-mentioned limitation for the density
matrix elements involved, translates into a closed system of equations
of motion for a reduced number of relevant expectation values. These
are
\begin{eqnarray}
 c_n &=& \left < \left|1\right> \left<1\right| \, b^{\dagger\,n}b^n
 \right>  \;, \qquad \qquad  n = 0,1,2 \ldots\; ,\nonumber \\   
 v_n &=& \left <  \left|2\right> \left<2\right| \, b^{\dagger\,n}b^n \right> \;, \qquad
 \qquad  n = 0,1,2 \ldots \quad \mathrm{and} \label{eq:cvt} \\
 t_n &=& -i g\left < \left|2\right> \left<1\right|\,  b^{\dagger\,n}b^{n-1} \right> \; ,
 \quad  \, n = 1,2,3 \ldots \nonumber
\end{eqnarray}
Obviously, the average population of the upper (lower) level is given by 
$c_0$ ($v_0$), which obey $c_0+v_0=1$.  Of a special interest for the
photon statistics 
are the expectation values $p_n =\left< b^{\dagger n}b^n \right>=
c_n+v_n$, and in particular the average photon number $p_1$, which
is also denoted by $N$.    
The imaginary prefactor in the definition of the multi-photon assisted
polarization $t_n$ makes it a real-valued quantity. 

The equation of motion for the expectation value of a given operator $A$
can be obtained from the master equation via
\begin{eqnarray}
\frac{\partial}{\partial t} \left<A\right> = \mathrm{Tr}\left \{A
\frac{\partial}{\partial t} \rho\right\} =  -i \left<\left[A, H_{JC}
  \right]\right>  
  &+&\frac{\kappa}{2}\left< \left[b^{\dagger},A\right]b +
  b^{\dagger}\left[A,b \right]\right > \nonumber \\
  &+& \sum_{(i,j)} \frac{\gamma_{ij}}{2}\left<
  \left[\sigma_{ij}^{\dagger},A\right]\sigma_{ij} + 
  \sigma_{ij}^{\dagger}\left[A,\sigma_{ij} \right]\right > \; .
\label{eq:eom}
\end{eqnarray}
For those in Eq.~(\ref{eq:cvt}) this leads to the following equations of motion
and the corresponding steady-state conditions \cite{agarwal,delvferm,pg} 
\begin{eqnarray}
\frac{\partial}{\partial t} c_n &=&
     -(n\kappa+\gamma)c_n + P v_n -2 t_{n+1} = 0 \;  ,\label{eq:eoma}\\
\frac{\partial}{\partial t} v_n &=&
     \gamma c_n -(n\kappa+P)v_n +2t_{n+1} +2nt_n = 0 \; , \label{eq:eomb}\\
\frac{\partial}{\partial t} t_n &=&
      g^2c_n + g^2 nc_{n-1} -g^2 v_n
      -\frac{(2n-1)\kappa+P+\gamma}{2}t_n = 0 \; .     
      \label{eq:eomc} 
\end{eqnarray}
By adding Eqs.~(\ref{eq:eoma}) and ~(\ref{eq:eomb}) one obtains the 
steady-state balance relation between the losses from the cavity and
its feeding through the photon-assisted polarization     
\begin{equation}
  \kappa \; p_n = 2 \; t_n \; , \quad n\geq 1 \; . \label{eq:balance} 
\end{equation}
Using this condition and Eq.~(\ref{eq:eoma}) with $n=0$, one obtains 
$P\,v_0=\gamma\, c_0 + \kappa\,N$, which allows to express the
steady-state level occupancies in terms of the photon output $N$
\begin{equation}
c_0 = \frac{P-\kappa\,N}{P+\gamma} \; , \quad
v_0 = \frac{\gamma+\kappa\,N}{P+\gamma} \; .
\label{eq:c0v0}
\end{equation}
 
The unknowns $c_n$ and
$v_n$ can be eliminated from Eqs.~(\ref{eq:eoma}) and
~(\ref{eq:eomb}) in favor of $t_n$ and, using again the balance
condition Eq.~(\ref{eq:balance}), one is lead\cite{agarwal} to a
three-term recursion equation for the photonic quantities $p_n$
\begin{equation}  
 A_n\,p_{n+1} + B_n\, p_n - C_n \, p_{n-1} = 0 \; , \quad n \geq
 1 \; , \label{eq:recrel} 
\end{equation}
with
\begin{eqnarray}
A_n &=& \frac{2\,\kappa}{n\, \kappa + P + \gamma} \; ,\nonumber \\
B_n &=& \frac{n\, \kappa -P+ \gamma}{n\, \kappa + P + \gamma}  +
        \frac{n\,\kappa}{(n-1)\,\kappa + P + \gamma} + \kappa \,
        \frac{(2n-1)\, \kappa + P 
          +\gamma}{4g^2} \; , \label{eq:ABC} \\ 
C_n &=& \frac{n\,P}{(n-1)\,\kappa + P + \gamma} \; \nonumber.  
\end{eqnarray}

By using the well-established connection between three-term recursion
problems and continued fractions\cite{Contfr}, Eq.~(\ref{eq:recrel})
allows for obtaining directly steady-state values, 
without resorting to the time evolution\cite{pg}. The convergence of
the continued-fraction solution is very good in all points, except the
intermediate pumping region where the transition takes place and the
population inversion becomes linear. 

In that interval an excellent agreement with the numerical solution can be
obtained by the following simple ansatz: Assume that (i)
Eq.~(\ref{eq:recrel}) is valid for $n=0$ too, 
and (ii) the last term $C_0\,p_{-1}$ takes in this case the value 0,
so that one has  
\begin{equation}  
 A_0\,p_1 + B_0\, p_0  = 0 \; .
\label{eq:ansatz} 
\end{equation} 
With $p_0=1$ and the values $A_0$ and $B_0$ as in
Eq.~(\ref{eq:ABC}) one obtains for the average photon number
\begin{equation} 
N = -\frac{B_0}{A_0} =  \frac{P-\gamma}{2\kappa} -
                      \frac{(P+\gamma)(P+\gamma-\kappa)}{8g^2} \; .  
\label{eq:ansatz_N} 
\end{equation}  
Using this in Eq.~(\ref{eq:c0v0}) leads for the population inversion 
$w=c_0-v_0$ to 
\begin{equation}  
w= \kappa \; \frac{P+\gamma -\kappa}{4g^2} \; ,
\label{eq:ansatz_w} 
\end{equation}
showing that the linear $P$-dependence of $w$ is a direct consequence of the
ansatz. 

In the scaling limit, $\kappa \to \varepsilon \kappa$, $g\to
\sqrt{\varepsilon} g$ with $ \varepsilon \to 0$,
 these results become
\begin{equation}  
w= \kappa \; \frac{P+\gamma}{4g^2} \; ,
\label{eq:scaled_ansatz_w} 
\end{equation}
for the population inversion and 
\begin{equation} 
\tilde N =  \frac{P-\gamma}{2\kappa} -
                      \frac{(P+\gamma)^2}{8g^2} \; .  
\label{eq:scaled_ansatz_N} 
\end{equation}  
for the rescaled photon output $\tilde N= \varepsilon N$.

Similar results can be obtained for the three-level model (details of
the calculations are left for a future publication)
with the conclusion that in the scaling limit the population inversion
behaves in as in Eq.~(\ref{eq:scaled_ansatz_w}) above, while 
the rescaled photon population obeys
\begin{equation} 
\tilde N =  \frac{(P-\gamma)\gamma'}{\kappa(P+2\gamma')} -
      \frac{(P+\gamma)(P\gamma+P\gamma'+\gamma\gamma')}{4g^2(P+2\gamma')} \; .  
\label{eq:scaled_ansatz_N_3} 
\end{equation}
Note that in the limit of large $\gamma'$, that is for very fast
$\left|3\right> \to \left|1\right>$ relaxation, one recovers the
two-level result, Eq.~(\ref{eq:scaled_ansatz_N}), as expected.
The agreement of these expressions,
Eqs.~(\ref{eq:scaled_ansatz_w}--\ref{eq:scaled_ansatz_N_3}), with the
numerical simulations is illustrated in Figs.~\ref{fig:two} and 
\ref{fig:four}. 

It is clear that the ansatz makes sense only in the interval of $P$
values for which $N \geq 0$. The first term in
Eq.~(\ref{eq:ansatz_N}) is positive if $P$ is not 
too small. On the other hand, for large $P$ values, the second,
negative term becomes dominant, so that the positivity condition can
hold only for a finite interval. 

Needless to say, the very good agreement between the  ansatz and the
numerical results (in the interval where the former makes sense) 
does not constitute a valid proof of the former. It is not immediately
obvious why Eq.~(\ref{eq:recrel}) should hold for $n=0$. Neither can one
use $C_0=0$ as an argument to replace the last term $C_0\,p_{-1}$ with
0\cite{delvmoll}, because it also contains the ill-defined $p_{-1}$. 
In order to prove the result one has to show first that, indeed, the
three-term recursion relation can be extended for $n=0$,
identifying in the process the quantity appearing in the role of
$C_0\,p_{-1}$. In a second step, one has to show that, in certain
conditions, this quantity does vanish, as required by the ansatz.
  
The first step of this program is the easier part. It relies on the
Glauber-Sudarshan (GS) $\cal{P}$-representation for the photonic density
operator \cite{Carmichael} as an integral over the complex plane of
coherent states
\begin{equation}  
 \rho =  \int \left| \alpha \right>  {\cal P}(\alpha) \left< \alpha
 \right| \, \frac{\mathrm{d}^2\alpha}{\pi} \; . 
\label{eq:GS} 
\end{equation}
Since there is no preferred phase angle in the theory (the reduced
photonic density operator, obtained by tracing out the emitter
indices, is diagonal in the photon number basis) the 
${\cal P}$-function depends only on $s = \left| \alpha \right|^2$,
${\cal P}(\alpha) = {\cal P}(s)$. Then, the normal-ordered photonic
expectation values $p_n$ turn out to be moments of the
quasi-distribution defined by  ${\cal P}$ 
\begin{equation}
p_n = \int_0^{\infty} s^n {\cal P}(s) \, \mathrm{d}s \, .
\label{eq:moments} 
\end{equation}
Note, for further reference, that the Poissonian statistics,
characterized by $p_n=\lambda^n$ 
would correspond to a sharp peak in the GS function 
${\cal P}(s)=\delta(s-\lambda)$ with $\lambda > 0$.
 
The ${\cal P}$-representation contains, in principle, the same
information as the density operator, but here we take advantage that
it allows a natural extension for the definition of expectation
values. A case in point is $p_n$ which, using Eq.~(\ref{eq:moments}),
becomes well-defined even for $n$ taking continuous, (not just
integer) positive values.  
Using the differential equations which translate the master equation
Eq.~(\ref{eq:master}) into the ${\cal P}$-representation formalism 
\cite{Carmichael}, one recovers the three-term recursion formula with a
continuous index, including the result for index zero\cite{pg}. The
latter can also be obtained by taking the limit  
$n \to 0$ in Eq~(\ref{eq:recrel}) and using for evaluating  
$C_n\,p_{n-1}$
\begin{equation} 
np_{n-1} = \int_0^{\infty} (s^n)' {\cal P}(s) \, \mathrm{d}s = 
         - \int_0^{\infty} s^n {\cal P'}(s) \, \mathrm{d}s
          \; \xrightarrow {n \to 0} \; {\cal  P}(0) \; .
\label{eq:nto0} 
\end{equation}
It is now obvious that the third term in the recursion relation
Eq.~(\ref{eq:recrel}) for $n=0$ is proportional to ${\cal P}(0)$ and
therefore the validity of the ansatz is equivalent to the requirement
that ${\cal P}(s)$ vanishes for $s=0$. 

The second task is to establish the conditions when the vanishing takes place.
This is the central point not only in the justification of the ansatz, but also
in the proof of the scaling limit result. The latter amounts to showing that in
this limit the ${\cal P}$-function becomes a $\delta$-distribution concentrated
on a positive value, and this obviously entails that indeed, its value at the
origin becomes vanishingly small.  

To this end we look at the differential equation obeyed by ${\cal
  P}(s)$, and which 
translates the master equation into the language of the 
${\cal P}$-representation. We summarize here the main steps, the details can be
found in \cite{pg}. To start with, the density matrix of our problem has a
two-by-two block structure, corresponding to the two levels of the emitter.
Correspondingly one has four ${\cal P}$-functions placed in a matrix ${\cal
P}_{i,j}(s), \; i,j = 1,2 $ and the photonic function we are interested in is
obtained by tracing out the emitter-state indices 
${\cal P}= {\cal P}_{1,1}+{\cal P}_{2,2}$. Using the rules for mapping the
master equation for $\rho$ into a Fokker-Planck equation for ${\cal P}$,
\cite{Carmichael} one is lead to a system of equations which is the counterpart
of Eqs.~(\ref{eq:eoma}--\ref{eq:eomc}). After eliminating ${\cal P}_{1,1}$ and 
${\cal P}_{2,2}$ in favor of ${\cal P}$ we obtain a second-order differential
equation for the latter. 

This equation has then to be analyzed in the scaling limit. To simplify the
notation we take $\kappa$ itself as the scaling parameter which goes to 0, and
impose the condition $g\to 0$ with $g^2/\kappa$ fixed, by writing 
$g^2 = \tilde g^2 \kappa$ and keeping $\tilde g^2$ constant. 
The rescaled expectation values $\tilde p_n= \kappa p_n $, which are the
object of the scaling statement, can be obtained as moments of a rescaled 
${\cal P}$-function
\begin{equation}
\tilde p_n = \int_0^\infty \kappa^n s^n {\cal P}(s) \, \mathrm{d}s
           = \int_0^\infty t^n \tilde {\cal P}(t) \, \mathrm{d}t \; ,
\label{eq:tilde_pn} 
\end{equation} 
with $t = \kappa s$ and 
\begin{equation}
   \tilde {\cal P}(t) = \frac{1}{\kappa}\, {\cal P}
              \left(\frac{t}{\kappa}\right) \; .
\label{eq:tilde_P}
\end{equation}     
If, indeed, the number of photons increases to infinity in the scaling limit,
then the ${\cal P}$ quasi-distribution function moves its weight to larger and
larger values and its moments cease to exist. In this situation only the
rescaled function remains meaningful. 
Intuitively, according to Eq.~(\ref{eq:tilde_P}), the graph of the
rescaled function $\tilde {\cal P}(t)$ is obtained from that of the
original ${\cal P}(s)$  by compressing the 
latter by a factor of $1/\kappa$ along the abscissa and expanding it
by the same factor along the ordinate. This would bring the rescaled
function to $\delta(t)$, in the limit $\kappa \to 0$, were it not for
the opposite tendency of ${\cal P}(s)$ to move away from the origin, as
discussed above. The net result of these competing trends is what one has to
establish. It is easy to rewrite the differential equation obeyed by 
${\cal P}(s)$ into the corresponding one for $\tilde {\cal P}(t)$, and retain
in the coefficients only the dominant terms in the scaling parameter $\kappa$.
The result is \cite{pg}:
 \begin{equation}
\frac{t^2}{4\tilde g^2}\, \kappa^2 \, \tilde {\cal P}''
      - \left[ \left(3\, \frac{\gamma+P}{8\tilde g^2} +1 \right) t 
        -\frac {1}{2}\, P \right] \, \kappa \, \tilde {\cal P}'
      + \left (t - \nu \right) \, \tilde {\cal P} = 0  \; ,
\label{eq:tilde_P_ode}
\end{equation}  
with $\nu$ an essential parameter in the discussion
\begin{equation}  
\nu = \frac{P-\gamma}{2} -
                       \frac{(P+\gamma)^2}{8 \tilde g^2} \; ,
\label{eq:nu} 
\end{equation}
whose $P$ dependence is important and therefore sometimes emphasized
by the notation $\nu(P)$.
Note that $\nu$ is the same as the rescaled photon population $\kappa
N$, see Eq.~(\ref{eq:scaled_ansatz_N}), so that they change sign
simultaneously.
   
The appearance of the small parameter $\kappa$ along with the derivatives
suggests a WKB approach to the $\kappa \to 0$ asymptotics of the
solution. In other words one searches the solution, up to a
normalization factor, in the form
\begin{equation} 
\tilde {\cal P}(t) = \mathrm{exp} \left (-\frac{1}{\kappa}\;
\varphi(t) \right) \; ,   
\label{eq:wkb} 
\end{equation}      
in which $\varphi(t)$ is taken in the leading, zeroth order in
$\kappa$. It is clear that when $\kappa$ gets smaller, the value of 
$\tilde {\cal P}(t)$ around the minimum of $\varphi(t)$ is greatly enhanced,
in comparison with the values at other points which, in the view of
normalization, become negligible. In the limit one obtains a
$\delta$-function concentrated at the minimum of $\varphi(t)$.   

The equation obeyed by  $\varphi(t)$ in the leading order has the form
of a quadratic equation for its derivative
\begin{equation} 
\frac{t^2}{4\tilde g^2}\, (\varphi')^2
      + \left[ \left(3\, \frac{\gamma+P}{8\tilde g^2} +1 \right) t 
        -\frac {1}{2}\, P \right] \, \varphi'
      + \left [t - \nu(P) \right] = 0  \; .
\label{eq:d_phi}
\end{equation}   
Around $t=0$ one of the roots behaves like $\varphi'\sim 2\tilde g^2
P/t^2$, i.e.  $\varphi \sim - 2\tilde g^2 P/t $  which in
Eq.~(\ref{eq:wkb}) leads to a strongly singular solution. The regular one
comes from the other root for which $\varphi'(0) = -2\nu(P)/P$.  

Two cases arise, depending on the sign of $\nu(P)$: (i) As long as
$\nu(P)$ is negative, $\varphi'(0) > 0$, then $t=0$ is a minimum for
$\varphi(t)$ and, according to the above discussion, 
$\tilde {\cal P}(t)$ tends to $\delta(t)$ 
in the scaling limit. (ii) When $\nu(P)$ becomes positive,
$\varphi'$ starts at $t=0$ with negative values and crosses the
abscissa at $t=\nu$. Then $\varphi(t)$ has a local maximum at the
origin and therefore the values of $\tilde {\cal P}(0)$ become
vanishingly small in the limit $\kappa \to 0$. This is precisely the
requirement for the ansatz to hold. The function
$\tilde {\cal P}(t)$ is now concentrated at the point of minimum
$t=\nu(P)$. 

As a consequence, in the interval in which $\nu(P)$ is positive one has 
$\tilde {\cal P}(t) \to \delta(t-\nu)$ and all the rescaled expectation
values become $\tilde p_n = \nu^n, n \ge 0$. Outside this interval
$\tilde {\cal P}(t) \to \delta(t)$ and all  $\tilde p_n = 0 $,
except $\tilde p_0=p_0$, which is equal to 1 by definition. The change
is abrupt and  takes place at the interval endpoints defined by the
quadratic equation $\nu(P) = 0$. The condition for this equation to
have real roots is $\tilde g^2 \geq 2\gamma $, or $g^2 \geq 2\kappa
\gamma $, and then the roots are both positive   
\begin{equation} 
 P_{\pm} = 2 \tilde g^2 -\gamma \pm 2 \tilde g \sqrt{\tilde g^2 -2\gamma} \; .
\label{eq:roots} 
\end{equation}
The lowest one, $P_{-} = P_{thr}$, corresponds to the onset of lasing
and the highest, $P_{+} = P_{sq}$, to self-quenching. The condition
$\tilde g^2 \geq 2\gamma$ distinguishes the two behaviors illustrated
in Fig.\ref{fig:one} because, when not fulfilled, no transition takes
place. With this, the proof of the scaling limit is complete.  

It is instructive to see the action of the scaling limit directly on
the recursion relation Eq.~(\ref{eq:recrel}). The essential point is
the observation that the
$n$-dependence of the coefficients $A_n, B_n, C_n$ gradually
disappears in the limit $\kappa \to 0$. More precisely, the recursion
for the rescaled expectation values  
\begin{equation}  
 \frac{A_n}{\kappa}\,\tilde p_{n+1} + B_n\, \tilde p_n - \kappa \,C_n
 \,\tilde p_{n-1} = 0 \; , \quad n \geq 1 \; ,  
\label{eq:renrecrel} 
\end{equation}
reduces, in the $\kappa \to 0$ limit, to
\begin{equation}  
\tilde p_{n+1} = -\kappa \,\frac{B_0}{A_0} \, \tilde p_n  = 
\nu \,\tilde p_n \; , \quad n \geq 1 \; ,   
\label{eq:twotermrecrel} 
\end{equation}  
with the obvious solution $\tilde p_n = \nu^{n-1} \tilde p_1$. 
For the values of the pump where $\nu(P)$ is negative only the trivial
solution $\tilde p_n =0, \; n \geq 1 $ is possible, in order to avoid
negative results for positive expectation values. On the other hand,
when $\nu(P)>0$, Eq.~(\ref{eq:twotermrecrel}) holds for $n=0$ too and
one has $\tilde p_1= \nu \, \tilde p_0 = \nu$. Then the solution is
exponential $\tilde p_n= \nu^n,\; n \ge 0$ in accordance with the
Poisson statistics. 

\begin{figure}
   \begin{center}
   \includegraphics[height=5cm]{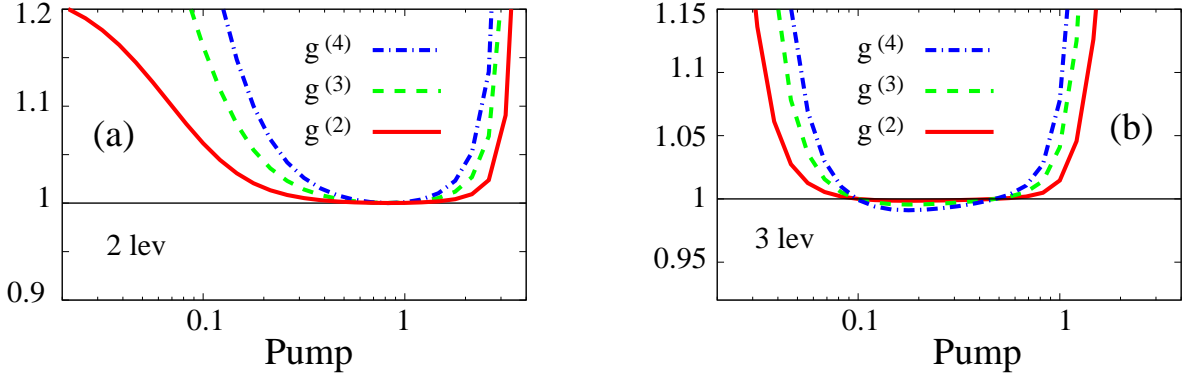}
   \end{center}
   \caption
   { \label{fig:five}
 Second, third and fourth correlation function for (a) the two-level
 emitter and (b) the three-level emitter. The parameters are the same
 as in Fig.\ref{fig:two} with $\varepsilon=1$ in (a) and
 $\varepsilon=0.004$ in (b). 
\vspace{0.3cm} 
 }
   \end{figure}

As  $\kappa$ approaches 0, the product  $n \kappa$  in the
coefficients of the recursion relation vanishes, and this is how their
$n$-dependence is lost. In the process it is the low-index
coefficients that are the first to get close to their limit values,
because the limit requires the product $n \kappa$ to be
small. Therefore the Poissonian condition $g^{(n)}=1$ is obeyed by
$g^{(2)}$ first, and by $g^{(3)},g^{(4)}, \dots$ only later. This is
numerically confirmed, as seen in Fig.\ref{fig:five}, and shows that
using $g^{(2)})=1$ as a criterion of truly coherent light may be, in
this sense, somewhat premature.

\section{CONCLUSION} \label{sec:conclusion}

By solving the master equation for a single emitter in JC
interaction with a cavity mode one observes a sudden change in the
behavior of the steady-state solution. This is indicative of the
onset of lasing and offers the possibility of identifying the
conditions for a sharp transition to a pure, as opposed to
approximate, Poissonian statistics. We have shown that these conditions
imply an asymptotic regime for the parameters controlling the generation
and loss of cavity photons. Specifically, the domain of parameters for
which a sharp transitions occurs is defined by both the cavity loss
$\kappa$ and the JC coupling $g$ going to 0, provided that $g$ scales
like $\sqrt{\kappa}$.  

The result is supported by numerical data, as exemplified for a two-level
and a three-level emitter, and is proven using analytical methods for the
two-level model. In a previous paper\cite{pg}, the same scaling limit
was shown to give rise to a sharp transition and to reproduce the
threshold value known in the literature, for the random injection model
of Scully and Lamb. It should be noted that the Scully-Lamb model does
not belong to the class considered here: while the latter 
are ``embedded'' JC systems the former is rather an ``intermittent'' JC
one. This fact, together with the numerical evidence, suggests that our
scaling limit result has a range of validity that is larger than the
set of cases for which a full analytic proof is available now.        
   

\bibliography{liter}  

\end{document}